# Dynamical model of DNA-protein interaction:

# effect of protein charge distribution and mechanical properties

Ana-Maria FLORESCU and Marc JOYEUX [(#)]

*Laboratoire de Spectrométrie Physique (CNRS UMR 5588),*

*Université Joseph Fourier - Grenoble 1,*

*BP 87, 38402 St Martin d'Hères, FRANCE*

**Abstract :** The mechanical model based on beads and springs, which we recently proposed to study non-specific DNA-protein interactions [J. Chem. Phys. 130, 015103 (2009)], was improved by describing proteins as sets of interconnected beads instead of single beads. In this paper, we first compare the results obtained with the updated model with those of the original one and then use it to investigate several aspects of the dynamics of DNA sampling, which could not be accounted for by the original model. These aspects include the effect on the speed of DNA sampling of the regularity and/or randomness of the protein charge distribution, the charge and location of the search site, and the shape and deformability of the protein. We also discuss the efficiency of facilitated diffusion, that is, the extent to which the combination of 1D sliding along the DNA and 3D diffusion in the cell can lead to faster sampling than pure 3D diffusion of the protein.

[(#)] email : marc.joyeux@spectro.ujf-grenoble.fr

# 1. Introduction

Although great advances have been made in genetics in the last decades and the genomes of several species are now completely mapped, there is still a lot of discussion on how gene expression takes place. This process is regulated by transcription factors, which are proteins that first connect to the DNA chain at specific sites and then promote transcription by RNA polymerase. The means by which these proteins find their targets are, however, not very well understood. Some of the first studies in this field suggested that the LacI repressor of *Escherichia coli* connects to its specific site at a rate that is much faster than could be expected by normal diffusion in the cell [1]. This triggered the development of a lot of theoretical models [2-8], which aimed at understanding the results of these experiments and at describing the mechanism of non-specific (sequence independent) as well as specific (sequence dependent) DNA-protein interactions. Most of these models rely on the hypothesis that the sampling of DNA is accelerated by a reduction of dimensionality [9], in the sense that part of the sampling is done in only one dimension, with the protein sliding along the DNA chain. This process is known as "facilitated diffusion": the protein connects randomly on the DNA chain and then slides along it in search of its target. If it does not find it after a certain amount of time it disconnects, diffuses in the cell and then reconnects somewhere else. Sampling may be greatly accelerated if random walks are interrupted before too many half turns, thereby diminishing the redundant part of the process.

The development of new techniques for *in vivo* microscopy has recently permitted the direct visualization of the motion of proteins inside the cell and the precise determination of their diffusion coefficients [10-19]. Nonetheless, the exact mechanism of non-specific DNA-protein interactions still remains unclear, because proteins come into a multitude of shapes and sizes, and each of them interacts with DNA in a particular manner. We recently proposed a dynamical model for the description of non-specific DNA-protein interaction, which we hope is sufficiently general to grab the traits that are common to all of these systems [20]. This model is based on the "beads and springs" description of the DNA chain, with elastic, bending and Debye-Hückel electrostatic interactions between the beads [21], while the protein interacts with DNA through electrostatic and excluded-volume forces. We studied the properties of this model using a Brownian dynamics algorithm that takes hydrodynamic interactions into account and obtained results that agree fairly well with the experimental results [10-19], as well as the assumptions and predictions of kinetic models [2-8]. For example, the protein samples DNA by a combination of three-dimensional diffusion (3D) in the cell and one-dimensional (1D) sliding along the DNA chain. This model evidences the presence, in a certain range of values of the effective protein charge, of facilitated diffusion, that is a combination of the two types of diffusion that leads to faster-than-3D diffusion sampling of DNA. Moreover, the analysis of sliding events showed that the number of base pairs visited during sliding increases with the square root of time and is comparable to those deduced from single molecule experiments. At last, the model suggests that, for the global (1D+3D) motion of the protein, this number increases linearly with time until it reaches a value that is close to the total number of DNA base pairs in the cell.

To our opinion, the weakest feature of this model is the over-simplified description of the protein, which was modeled as a single rigid sphere [20]. The purpose of the present work is to update this model with a more decent description of the protein and to check to what extent this affects the conclusions drawn in [20]. More precisely, we model the protein as a set of 13 beads, which carry different electrostatic charges and are interconnected by springs, and discuss whether the improved model still displays facilitated diffusion. We also study how different parameters - like the total protein charge, the charge and location of the search site, the randomness of the charge distribution, and the shape and the deformability of the protein – affect the efficiency of the sampling process.

The remainder of this article is organized as follows. The improved model is described in Sect. 2, paying special attention to the geometries and charge distributions of the various protein configurations which dynamics are studied in this paper. Sect. 3 reinterprets to some extents the results obtained in [20] in the light of what is known as the "volume of the Wiener sausage" in the mathematical literature, while the results obtained with the improved model are presented in Sect. 4 to 6. More precisely, the laws governing the time evolution of the number of different DNA beads visited by the protein search site during 1D sliding and 3D motion in the cell are discussed in Sect. 4. The relevance of the concept of facilitated diffusion for the new model is next analyzed in some detail in Sect. 5. Sect. 6 finally describes the effect on the speed of DNA sampling of several protein properties that could not be taken into account within the single bead protein model of [20], like the shape and deformability of the protein, the regularity or randomness of its charge distribution, and the charge and position of the search site. We conclude in Sect. 7.

## 2 - Model and simulations

The system consists of DNA and a protein enclosed in a sphere, which models the cell or its nucleus. As discussed in [20], DNA is not modeled as a single long chain, in order to avoid excessive DNA curvature at the cell walls. It is instead modeled as a set of $m$=50 disconnected smaller chains (hereafter called *segments*), each segment consisting of $n$=40 beads, which are separated at equilibrium by a distance $l_0 = 5.0$ nm. Each bead, which represents 15 base pairs, has a hydrodynamic radius $a_{\text{DNA}} = 1.78$ nm and an effective charge $e_{DNA} = -0.243 \times 10^{10} l_0 \bar{e} \approx -12 \bar{e}$ placed at its centre ($\bar{e}$ is the absolute charge of the electron). The radius of the sphere, $R_0 = 0.169$ μm, was chosen in order that the density of bases inside the cell is close to the experimentally observed one. Indeed, as pointed out in [5], the volume $V$ of the cell or the nucleus is connected to the total DNA length $L$ according to $V = w^2 L$, where $w$ represents roughly the spacing of nearby DNA segments. $m$ and $n$ must therefore fulfil the relation $\frac{4}{3}\pi R_0^3 \approx w^2 mn l_0$, where the average value $w = 45.0$ nm holds for both prokaryote and eukaryote cells. Moreover, the length of each DNA segment is approximately equal to the radius of the cell, that is $nl_0 \approx R_0$, so that (i) the cell is rather homogeneously filled with DNA, (ii) end effects are negligible, and (iii) excessive curvature of DNA segments touching the cell wall is avoided.

While proteins were taken as single beads in [20], they are modeled in the present work by sets of 13 beads with hydrodynamic radius $a_{prot} = 3.5$ nm interconnected by elastic bonds. We essentially investigated two different protein geometries, namely "spherical" and "linear" proteins. The simplest "spherical" protein is obtained by placing 12 beads at the vertices of a regular icosahedron and a thirteenth bead at its center (21 beads would have been required for a regular dodecahedron). A bond connects the central bead to the 12 other beads, and each bead at a vertex is connected to its five nearest neighbors by a similar bond. The distance between the central bead and those at the vertices is equal to the bead radius $a_{prot} = 3.5$ nm, so that the radius of the protein at rest is close to 7.0 nm and the distance between two nearest neighbors placed at vertices is $L_0 = 4a_{prot}/\sqrt{10 + 2\sqrt{5}} \approx 3.68$ nm. Linear proteins are taken as flexible and extensible chains of 13 beads separated at equilibrium by a distance $a_{prot} = 3.5$ nm. Because no bending interaction among protein beads is taken into account (see below), "linear" proteins generally assume bent geometries with average end-to-end distances of the order of 17.0 nm. We fixed the number of beads of linear proteins to 13 for the sake of an easier comparison with spherical proteins.

All beads, except for those at the center of spherical proteins, are assigned electrostatic charges $e_p$ placed at their centers (note, however, that electrostatic interactions between protein beads are neglected, see below). We considered several protein charge distributions, namely (i) uniform distributions with increasing total charge $e_{\text{prot}} = \sum_p e_p$, (ii) gradients of charges with fixed total charge $e_{\text{prot}}$ and increasing values of the maximum charge $e_{\max}$, (iii) gradients of charges with fixed maximum charge $e_{\max}$ and increasing total charge $e_{\text{prot}}$, and (iv) random distributions. For spherical proteins, gradient distributions are based on sets of four equally spaced charge values $e_{\max} - k\Delta$, where $k$ varies from 0 to 3 and $\Delta = (12 e_{\max} - e_{\text{prot}})/18$. Charges $e_{\max}$ and $e_{\max} - 3\Delta$ are carried by two beads placed at opposite vertices of the icosahedron, while the five beads closest to the bead with charge $e_{\max}$ carry a charge $e_{\max} - \Delta$ and the five beads closest to the bead with charge $e_{\max} - 3\Delta$ carry a charge $e_{\max} - 2\Delta$. For linear proteins, instead of a single bead with charge $e_{\max} - 3\Delta$, we placed the charge $(e_{\max} - 3\Delta)/2$ at the center of two beads, in order to compensate for the fact the bead placed at the centre of the icosahedron is not charged. We usually increased the total charge $e_{\text{prot}}$, as well as the maximum charge $e_{\max}$, up to $-5 e_{\text{DNA}}$, that is approximately $60\,\overline{e}$, which covers a wide range of experimental values for both prokaryote and eukaryote transcription factors [22,23]. At last, all the results presented below were obtained by considering that the protein search site corresponds to a single bead. In most cases, and unless otherwise specified, this bead had the highest positive charge $e_{\max}$, but we also ran simulations were this was no longer the case.

The potential energy $E_{\text{pot}}$ of the system is composed of four terms

$$E_{pot} = V_{DNA} + V_{prot} + V_{DNA/prot} + V_{wall} ,\qquad(2.1)$$

where $V_{DNA}$ describes the potential energy of the DNA segments and the interactions between them, $V_{prot}$ refers to interactions among protein beads, $V_{DNA/prot}$ stands for the interactions between the protein and the elements of the DNA chain, and $V_{wall}$ models the interactions with the cell wall, which maintain the protein and the DNA inside the cell (confinement energy). We used for $V_{DNA}$ the same expression as in our previous work (see eq. (2.2) of [20] and the discussion below this equation). $V_{DNA}$ is therefore modelled as the sum of stretching, bending and repulsive electrostatic terms. In contrast, we considered that the beads that compose proteins interact with each other only by means of harmonic stretching potentials. More precisely, for linear proteins

$$V_{prot} = \frac{1}{2} C \frac{k_B T}{a_{prot}^2} \sum_{j=0}^{11} \left(L_{j,j+1} - a_{prot}\right)^2 .\qquad(2.2)$$

In eq. (2.2), the 13 beads are labeled from $j=0$ to $j=12$ and $L_{j,j+1} = \|\mathbf{R}_j - \mathbf{R}_{j+1}\|$ denotes the distance between two successive beads ($\mathbf{R}_j$ is the position of bead $j$). A distance $a_{prot}$ separates two neighbouring beads at equilibrium. For spherical proteins, we instead assumed that

$$V_{prot} = \frac{1}{2} C \frac{k_B T}{a_{prot}^2} \sum_{j=1}^{12} \left(L_{0,j} - a_{prot}\right)^2 + \frac{1}{2} C \frac{k_B T}{L_0^2} \sum_{j=1}^{12} \sum_{\substack{k \in V_1(j) \\ k>j}} \left(L_{j,k} - L_0\right)^2 .\qquad(2.3)$$

In eq. (2.3), index 0 refers to the bead located at the center of the icosahedron and indices 1 to 12 to those placed at the vertices, $L_{j,k} = \|\mathbf{R}_j - \mathbf{R}_k\|$ denotes the distance between protein beads $j$ and $k$, and $k \in V_1(j)$ means that the sum runs over the five beads $k$ that are the nearest neighbours of bead $j$ at equilibrium. At equilibrium, the central bead is separated by $a_{prot}$ from the beads placed at the vertices of the icosahedron, while two neighbouring beads located at vertices are separated by $L_0$. Unless otherwise specified, all the results shown below were obtained with a constant $C$ in eqs. (2.2) and (2.3) equal to $C=100$, like for the DNA elasticity constant $h$, in order to get very small displacements of the average bond length without precluding the use of sufficiently large time steps. Still, we also ran simulations where $C$ was varied between 5 and 225 to study how the deformability of proteins affects facilitated diffusion.

The confinement potential $V_{\text{wall}}$ is taken, as in [20], as a sum of repulsive terms that act on the beads that trespass the radius $R_0$ and repel them inside the sphere:

$$V_{\text{wall}} = k_B T \sum_{j=1}^{m} \sum_{k=1}^{n} f\left(\|\mathbf{r}_{j,k}\|\right) + 10 \, k_B T \sum_{j=0}^{12} f\left(\|\mathbf{R}_j\|\right), \tag{2.4}$$

where $\mathbf{r}_{j,k}$ denotes the position of bead $k$ of DNA segment $j$ and $f(x)$ is the repulsive wall function defined in eq. (2.5) of [20]. Finally, and most importantly, the interaction between a protein and a DNA bead is described as the sum of a repulsive excluded volume term and an electrostatic Debye-Hückel potential, which might be either repulsive or attractive, depending on the sign of the electrostatic charge $e_p$ placed at the center of the protein

$$V_{\text{DNA/prot}} = \sum_{p=0}^{12} \left( E_e^{(p)} + E_{\text{ev}}^{(p)} \right)$$

$$E_e^{(p)} = \frac{e_{\text{DNA}} e_p}{4\pi\varepsilon} \sum_{j=1}^{m} \sum_{k=1}^{n} \frac{\exp\left(-\dfrac{1}{r_D} \|\mathbf{r}_{j,k} - \mathbf{R}_p\|\right)}{\|\mathbf{r}_{j,k} - \mathbf{R}_p\|} \tag{2.5}$$

$$E_{\text{ev}}^{(p)} = 1.86 \, k_B T \left|\frac{e_p}{e_{\text{DNA}}}\right| \sum_{j=1}^{m} \sum_{k=1}^{n} F\left(\|\mathbf{r}_{j,k} - \mathbf{R}_p\|\right) ,$$

where the function $F(x)$, which is defined in eq. (2.7) of [20], is the repulsive part of a Lennard-Jones-like potential function. Note that charges are taken as signed quantities in eq. (2.5), while they were considered as positive quantities in eq. (2.6) of [20]. This is the reason why there was a minus sign in the expression for $E_{\text{ev}}^{(p)}$ in our first study, while there is none in the present work. The reader can refer to [20] for a discussion concerning the choice of the excluded volume potential $E_{ev}^{(p)}$. However, it is important to emphasize that, when the charges placed at the centre of the DNA and protein beads have opposite signs, the interaction between the two beads must be minimum at some value close to $\sigma = a_{DNA} + a_{prot} = 5.28$ nm, i.e. the sum of the radii of the DNA and protein beads, in order for 1D sliding to take place. The expression for $E_{ev}^{(p)}$ in eq (2.5) insures that this is indeed the case and that the position of the minimum does not depend on the charge $e_p$. It should however be mentioned that, in this study, the interaction potential is minimum not when the centers of the two beads are separated by exactly σ, as in [20], but rather when this distance is equal to σ+0.5 nm (this was achieved by introducing the factor 1.86 in the expression of $E_{ev}^{(p)}$). The minimum of the potential well was shifted by this small amount in order to better agree with recent theoretical models [26] and experimental results for complexes of EcoRV [27] and the Skn1 and Sap1 proteins [28].

As in [20], we use the Brownian dynamics algorithm of Ermack and McCammon [29] to integrate numerically the equations of motion for the 13 protein beads and the 100 DNA beads that are located closest to them (see eq. (2.8) of [20]). The hydrodynamic interactions tensor $\mathbf{D}^{(n)}$, on which this algorithm is based, is built using a modified form of the Rotne-Prager hydrodynamic interaction tensor [30] for beads of unequal sizes [31-33] (see eqs. (26)-(28) of [33]). The CPU time required for the Choleski factorization of $\mathbf{D}^{(n)}$ in the Ermack and McCammon algorithm increases as the cube of the number of beads that are taken into account in $\mathbf{D}^{(n)}$, so that this turns out to be the limiting step for the investigation of the dynamics of large systems. This is the reason why we use the diagonal approximation of this algorithm, that is eq. (2.10) of [20], to compute the motion of the 1900 other DNA beads. As discussed in [20], this is sufficient to insure that Choleski factorization slows down calculations by only 10% without affecting the results of the present simulations, which essentially aim at studying the interactions between DNA and the protein. This procedure is therefore an interesting alternative to Fixman's approximation [34-36].

At last, let us mention that we checked that simulations performed with respective time steps $\Delta t$ of 25, 100 and 200 ps lead to similar results. All the results discussed below were consequently obtained with a time step $\Delta t = 100$ ps. Moreover, all these results were averaged over six trajectories with different initial configurations.

## 3 - 1D and 3D Wiener sausages

The purpose of this section is to re-interpret to some extent the results presented in our previous study [20] in the light of what is known as the "volume of the Wiener sausage" in the mathematical literature.

We showed in [20] that, when the protein is modeled by a single bead, the portion of time $\rho_{1D}$ during which it interacts with the DNA chain increases sharply and continuously with the electrostatic charge $e_{prot}$ placed at its center (see Fig. 4 of [20]). We also showed that the number $N(t)$ of different DNA beads visited by the protein after some given amount of time $t$ also increases with $e_{prot}$ up to about $e_{prot} \approx |e_{DNA}|$. However, it then remains approximately constant up to $e_{prot} \approx 3|e_{DNA}|$, before decreasing again rapidly for larger values of $e_{prot}$ (see Fig. 9 of [20]). The maximum speed up of the search time due to facilitated diffusion was found to be of the order of 2. Moreover, we observed that, for $e_{prot} \approx |e_{DNA}|$, $N(t)$ increases as the square root of time during 1D sliding events (see Fig. 8 of [20]). In contrast, when considering the global motion (1D+3D) of the protein, $N(t)$ was found to increase linearly with time (as long as it remains small compared to the total number of DNA beads) for the whole range of investigated values of $e_{prot}$ (see Fig. 13 of [20]). We concluded from these observations, that the model predicts that 1D sliding is a diffusive process, while 3D motion of the protein in the cell or its nucleus is not. It turns out that the second conclusion is erroneous, in the sense that a linear increase of $N(t)$ during 3D motion is actually characteristic of a diffusive behaviour. Indeed, if we assume that both the 1D and 3D motions of the protein are pure Brownian processes characterized by diffusion coefficients $D_{1D}$ and $D_{3D}$, such that

$$\langle \mathbf{R}^2 \rangle(t) = 2 D_{1D} t \tag{3.1}$$

for 1D motion, and

$$\langle \mathbf{R}^2 \rangle(t) = 6 D_{3D} t \tag{3.2}$$

for 3D motion, then the number $N(t)$ of different DNA beads visited by the protein after time $t$ is closely related to the length $\ell(t)$ (1D motion) or the volume $V(t)$ (3D motion) visited by the bead after the same time. These quantities were called the "volume of the Wiener sausage" by Donsker and Srinivasa [37]. Analytical expressions for their long time asymptotics are known [38,39]. For 1D motion, one has

$$\ell(t) \approx \sqrt{\frac{16}{\pi} D_{1D} t} \; , \tag{3.3}$$

while for 3D motion

$$V(t) \approx 4\pi \, \delta \, D_{3D} \, t, \tag{3.4}$$

where $\delta$ is the maximum distance away from the central Brownian motion. Most importantly, $V(t)$ increases linearly with time, which implies that $N(t)$ also does, because $N(t)$ and $V(t)$ can be related with good approximation through $N(t) = c V(t)$, where $c$ is the concentration of DNA beads inside the cell (this expression is based on the assumption that DNA motion can be neglected compared to the protein one). Therefore, the linear increase of $N(t)$ observed in [20] indicates that our model predicts that the 3D motion of the protein is essentially diffusive, just like 1D sliding, which agrees with the basic assumption of kinetic models. One can use eq. (3.4) to estimate the rate of linear increase $\kappa$ of $N(t)$

$$N(t) = \kappa\, t \tag{3.5}$$

as long as it remains sufficiently small compared to the total number of DNA beads. In our model it is sufficient that the protein bead touches a DNA bead to interact with it. Parameter $\delta$ in eq. (3.4) must therefore be taken as $\delta = a_{prot} + 2a_{DNA}$. By combining these equations, one therefore gets

$$\kappa \approx 4\pi \left(a_{prot} + 2a_{DNA}\right) D_{3D}\, c\ . \tag{3.6}$$

When plugging in eq. (3.6) the actual concentration of DNA beads, $c = 9.89 \times 10^{22}$ beads/m$^3$, and the 3D diffusion coefficient at 298 K of a sphere of radius $a_{prot}$,

$$D_{3D} = \frac{k_B T}{6\pi\, \eta\, a_{prot}} \approx 0.7 \times 10^{-10}\ \mathrm{m^2/s}\ , \tag{3.7}$$

one obtains $\kappa \approx 0.61$ beads/µs, which is less than a factor 2 away from the value $\kappa \approx 1.09$ beads/µs we observed for a purely repulsive potential between the protein and the DNA beads [20], and coincides almost perfectly with the value which is obtained when hydrodynamic interactions are furthermore disregarded, that is, when the positions of all the beads are updated according to eq. (2.10) of [20].

Similarly, one can use eq. (3.3) to estimate the diffusion coefficient of the protein during 1D sliding events along the DNA chain. We indeed observed that for $e_{prot} = |e_{DNA}|$ and for sufficiently long sliding events, $N(t)$ increases according to $N(t) = \alpha \sqrt{t}$ where $\alpha \approx 6.34$ beads µs$^{-1/2}$ (see Fig. 8 of [20]). Since $N(t) = \ell(t)/l_0$, one consequently obtains

$$D_{1D} = \frac{\pi}{16} l_0^2 \alpha^2 \approx 2.0 \times 10^{-10} \text{ m}^2/\text{s} , \qquad (3.8)$$

or, equivalently, $D_{1D} \approx 7.9$ beads$^2$/µs, or $D_{1D} \approx 1800$ bp$^2$/µs, which is, as already discussed in [20], about two orders of magnitude too large compared to experimental values [11,14].

**4 – Time evolution of $N(t)$ during 1D and 3D motion**

As discussed in some detail in Sect. 3, for single bead proteins $N(t)$ increases like $\sqrt{t}$ during 1D sliding events and like $t$ for the global 3D motion at short times, that is as long as $N(t)$ remains small compared to the total number of DNA beads. Is this still the case for 13 beads proteins ?

We investigated a large number of different spherical and linear 13 beads protein models and found that, for all the examples with reasonable charge distributions, $N(t)$ follows the law we proposed in [20] for single bead proteins, that is

$$\frac{N(t)}{2000} = 1 - \exp(-\kappa \frac{t}{2000}) . \qquad (4.1)$$

This is illustrated in Fig. 1, which shows the time evolution of $\log(1 - N(t)/2000)$ for selected linear and spherical proteins with uniform and gradient distributions of charges. It is seen that eq. (4.1) remains valid for very long times and for values of $N(t)$ very close to the total number of DNA beads. Eq. (4.1) of course reduces to eq. (3.5) at short times. According to the formula for the volume of the Wiener sausage, this indicates that, like for single bead proteins, the global motion of 13 beads proteins is essentially diffusive. Fig. 1 also points towards a very general result, namely that $N(t)$ increases significantly more rapidly for linear proteins than for spherical ones (at least as long as the search site is located at one of the extremities of the chain, see Sect. 6). The rationale for this observation is that, according to eq. (3.6), κ increases linearly with $D_{3D}$ and the 3D diffusion coefficient of linear proteins is significantly larger than that of spherical ones ($0.35 \times 10^{-10}$ against $0.20 \times 10^{-10}$ m$^2$/s at $C=100$).

In contrast, it might seem at first sight that 13 beads proteins differ more substantially from single bead ones as far as 1D sliding along DNA is involved. For example, Fig. 2 shows log-log plots of $N(t)$ for long sliding events of spherical proteins with uniform and gradient distributions of charges. It is seen that the time evolution of $N(t)$ approximately corresponds to straight lines in these plots, which implies that $N(t)$ increases as a power of $t$, that is $N(t) = \alpha \, t^{\beta}$, but the exponent β is now smaller than 1/2. Stated in other words, 1D sliding is subdiffusive. This is not really surprising, because subdiffusion is often encountered in dense media and has recently been experimentally reported for the *global* motion of proteins in the cytoplasm or the nucleus [40-42]. By looking more closely at sliding events, it can be noticed that 13 beads proteins spend large amounts of time attached to the same DNA bead and the time intervals during which they actually slide are substantially shorter than for single bead proteins with $e_{prot} = -e_{DNA}$. This is an important observation, because it is well known that large average waiting times between random-walk steps are sufficient to induce subdiffusion (see for example [43]). The reason why waiting times are longer for 13 beads proteins than for single bead ones is that in this model sliding is driven uniquely by thermal noise and this process is less efficient for 13 beads proteins than for single bead ones, because part of the energy received from collisions is used to deform proteins instead of being converted into sliding impulsions. Still, it should be mentioned that the average number of beads visited during each sliding event (5 to 10 beads, that is from 75 to 150 base pairs) is in fairly good agreement with experimental results, which lie in the range 30 to 170 base pairs [18,19].

If the depth of the attractive well between DNA and the protein is smaller than the energy $k_B T$ of thermal noise, then the protein does not spend enough time connected to DNA for actual sliding to take place. On the other hand, if attraction is too strong, then the protein remains attached to a particular DNA bead instead of sliding. One therefore expects that waiting times become longer and longer for increasing values of the protein charge $e_{prot}$ and, consequently, that the exponent β decreases. It can be checked in the top plot of Fig. 2 that this is indeed the case. While values of β close to 0.40 were obtained for most of the investigated proteins (see Fig. 2), β was found to decrease down to about 0.20 for uniform charge distributions with $e_{prot} = -4.8 e_{DNA}$ (see the top plot of Fig. 2).

At that point, we however checked that single bead proteins actually behave just like 13 beads ones with that respect. More precisely, we performed simulations with single bead proteins with charge $e_{prot} = -5 e_{DNA}$ and obtained $\beta \approx 0.30$. The diffusive character of 1D sliding reported in our previous work ($\beta = 0.50$ for $e_{prot} = -e_{DNA}$) therefore does not extend to proteins with too large values of $e_{prot}$.

To summarize this section, the time evolution of $N(t)$ is rather similar for single bead and 13 beads proteins. For the global 3D motion, $N(t)$ evolves according to eq. (4.1), which reduces to the linear law of eq. (3.5) at short times. The expression for the volume of the 3D Wiener sausage (eq. (3.4)) and the expression for κ derived there from (eq. (3.6)) therefore apply to both models. For moderate protein charges, the motion during 1D sliding is diffusive for single bead proteins ($\beta = 0.5$) and slightly subdiffusive for 13 beads proteins ($\beta \approx 0.40$). The subdiffusive character of the motion increases with $e_{prot}$ for both models. Strictly speaking, the expression for the length of the 1D Wiener sausage (eq. (3.3)) applies only to diffusive motion.

## 5 – Facilitated diffusion and speed up of DNA sampling

In [20], the value of the electrostatic charge placed at the center of the protein bead was increased in order to vary the portion of time $\rho_{1D}$ during which the protein is attached to DNA and check whether certain combinations of 1D and 3D motions lead to faster DNA sampling than pure 3D diffusion. In this section, we will follow the same general idea, except that the protein is now modeled by 13 interconnected beads instead of a single one, so that there are several different ways to modify $\rho_{1D}$.

The most natural way to compare the dynamics of the present model to that of the previous one consists in placing identical electrostatic charges at the centre of the 12 beads located at the vertices of the icosahedron (uniform charge distributions) and letting these charges vary. Results for such spherical proteins with uniform charge distributions are presented in Fig. 3. This figure displays the evolution, as a function of the total protein charge $e_{\text{prot}}$, of three quantities, namely $N(100\,\mu s)$, the number of different DNA beads visited by the protein search site after 100 µs (top plot), $\rho_{1D}$, the portion of time that the protein search site spends attached to a DNA bead (middle plot), and $n_{\text{sim}}$, the average number of DNA beads that are simultaneously attached to the protein search site when it interacts with DNA. Circles correspond to results obtained by considering that protein bead $p$ is attached to bead $k$ of DNA segment $j$ if $\|\mathbf{r}_{j,k} - \mathbf{R}_p\| \leq \sigma$, while lozenges correspond to the criterion $\|\mathbf{r}_{j,k} - \mathbf{R}_p\| \leq 1.5\,\sigma$. Error bars indicate the standard deviations for the six trajectories over which each point was averaged. The point at $e_{\text{prot}} = 0$ corresponds to purely repulsive DNA-protein interactions, that is, more precisely, when keeping only the repulsive part of the interaction potential with $e_p = -0.1 e_{\text{DNA}}$. It can safely be considered that, for repulsive DNA-protein interactions, the motion of the protein inside the cell is rather similar to pure 3D diffusion.

Examination of the middle and bottom plots of Fig. 3 shows that both $\rho_{1D}$ and $n_{sim}$ increase with $e_{prot}$ like for single bead proteins. Large values of $n_{sim}$ indicate that the protein charge is sufficiently large for the protein to attract and attach simultaneously to several DNA segments, which form a cage around it. As in [20], it is emphasized, that this phenomenon is probably not relevant from the biological point of view, because only a few proteins are known to have more than one "reading head" [44] (an example is precisely the Lac repressor, which has two binding sites [45]). This implies that one should consider only those charge distributions, which are associated with moderate values of $n_{sim}$, say, smaller than 3 for the $1.5\sigma$ threshold.

When comparing the top plot of Fig. 3 to Fig. 9 of [20], one first notices that $N(t)$ increases more slowly for 13 beads proteins than for single bead ones. For example, for the repulsive potential, the number of DNA beads visited by 13 beads proteins is only about 50% of the number of DNA beads visited by single bead proteins. This is again essentially due to the difference in the values of the 3D diffusion coefficient at 298 K, which is equal to $D_{3D} = 0.70 \times 10^{-10}$ m$^2$/s for single beads and to $D_{3D} \approx 0.20 \times 10^{-10}$ m$^2$/s for 13 interconnected beads. Nonetheless, the key point is certainly that, as for single bead proteins, there exists a range of values of $e_{prot}$ for which $N(t)$ increases more rapidly than for repulsive DNA-protein interactions. This range extends roughly up to $e_{prot} = -2e_{DNA}$. Still, it can be noticed that $N(t)$ is increased at maximum by about 50% compared to the repulsive potential, while a maximum increase close to 70% was obtained for single bead proteins [20]. This is probably connected to the point, already discussed in the previous section, that 1D sliding is slower and less efficient for 13 beads proteins than for single bead ones.

Needless to say that these conclusions drawn from the dynamics of proteins with uniform charge distributions must be confirmed by results obtained for more complex distributions. We postpone the case of random charge distributions till the next section and focus now on the results obtained for spherical proteins with gradient distributions of charges. For such gradient distributions, we either fixed the value of the maximum protein charge $e_{max}$ and varied the total charge $e_{prot}$, or fixed $e_{prot}$ and varied $e_{max}$. It turns out that the results obtained for these gradient distributions are quite similar to those discussed above, at least as long as $e_{prot}$ and $e_{max}$ remain moderate. For example, the results for $e_{max} = -0.8 e_{DNA}$ are shown in Fig. 4 and those for $e_{prot} = 0$ in Fig. 5. It is seen that, in both cases, $\rho_{1D}$ increases with increasing charge and $N(100\,\mu s)$ goes through a maximum for values of $\rho_{1D}$ comprised between 0.3 and 0.7 for the $1.5\sigma$ threshold. Moreover, the increase of $N(t)$ relative to the case of purely repulsive interactions between DNA and the protein does not exceed 40%, which again agrees with the results obtained for uniform charge distributions.

Things are however noticeably different for larger values of $e_{max}$ or $e_{prot}$. For example, we checked that for gradient distributions with $e_{prot} = -2.4 e_{DNA}$, the total protein charge is sufficiently large for proteins to spend all the time attached to a DNA segment, irrespective of $e_{max}$ (and consequently of the charge of the search site: we assumed so far that the search site is the protein bead with highest positive charge). As a consequence, $N(100\,\mu s)$ varies little with increasing values of $e_{max}$.

Conclusion therefore is that, even for rather rigid spherical protein models (remember that $C=100$ for all the results presented above), facilitated diffusion increases

DNA sampling speed by about 20 to 50% compared to 3D diffusion, which is substantially less than the 70% increase observed for single bead proteins [20]. Still, the efficiency of the facilitated diffusion mechanism is even lower for linear proteins, as can be seen in Fig. 6, which shows results obtained for linear proteins with uniform charge distributions (similar results were obtained for gradient distributions with $e_{prot} = 0$). $C$ was also fixed to 100. Since no clear increase of $N(100 \, \mu s)$ is observed when the total charge is increased from zero, in spite of the fact that $\rho_{1D}$ does increase significantly, it must be admitted that no combination of 1D and 3D motions is more efficient than pure 3D diffusion. This can be understood by noticing that, for identical values of $C$, spherical proteins are much more rigid than linear ones, because each bead at the vertices of the isocahedron is connected to the central bead and to its five nearest neighbours, while each bead of linear proteins is connected to only one or two nearest neighbours. 1D sliding of linear proteins is therefore still less efficient than that of spherical ones.

**6 – Which other parameters do affect the speed of DNA sampling ?**

The purpose of this section is to discuss the effect of several parameters, namely the value of the elastic constant $C$, the randomness of the charge distribution and the charge and position of the protein search site, on the speed of DNA sampling.

Let us first consider the effect of the protein elastic constant $C$. The time evolution of the number $N(t)$ of different DNA beads visited by the protein search site for spherical proteins with a gradient distribution of charges with $e_{prot} = -0.8 e_{DNA}$ and $e_{max} = -1.2 e_{DNA}$ and values of $C$ ranging from 10 to 225 is shown in the bottom plot of Fig. 2 for long sliding events and in Fig. 7 for the global (1D+3D) motion. While 1D sliding depends little on $C$, for the global motion $N(t)$ instead decreases significantly and rapidly with $C$ for values of *C* comprised between 10 and 100 before remaining nearly constant for larger values of *C*. It can be checked in Fig. 8 that this is essentially due to the evolution with increasing values of *C* of the diffusion coefficient, in agreement with eq. (3.6). The top plot indeed shows that $D_{3D}$ decreases from about $0.32 \times 10^{-10}$ m$^2$/s for *C*=5 to about $0.20 \times 10^{-10}$ m$^2$/s for values of C larger than 100. These values of $D_{3D}$ were obtained from eq. (3.2) by launching simulations that took only the protein into account and disregarded both DNA segments and cell boundaries. The average protein radius $\langle L_{0,j} \rangle$ was also computed during these simulations. Results are shown as filled circles in the bottom plot of Fig. 8. It is seen that $\langle L_{0,j} \rangle$ *decreases* with increasing values of *C* in the range *C*=5-100, so that the increase of $D_{3D}$ in this range is not in contradiction with eq. (3.7). This decrease of $\langle L_{0,j} \rangle$ with increasing *C* is rather counter-intuitive but agrees with preceding work [32]. It is actually due to hydrodynamic interactions. Indeed, if hydrodynamic interactions are neglected in the simulations aimed at estimating $D_{3D}$, then the more intuitive result that $\langle L_{0,j} \rangle$ slightly *increases* with *C* in the range *C*=5-100 is recovered (see the empty squares in the bottom plot of Fig. 8). Note that Kirkwood formula for estimating the equivalent hydrodynamic radius of the protein [46] totally fails to reproduce this behaviour. Conclusion therefore is that, like the shape discussed in Sect. 4, the deformability of the protein essentially affects the speed of DNA sampling through the associated variations of the diffusion coefficient.

Let us next turn to the effect of the regularity/randomness of the protein charge distribution. While all results presented up to now involved proteins with either uniform or gradient distributions of charges, Fig. 9 indicates how these results are affected when the charges of a gradient distribution are redistributed randomly. More precisely, Fig. 9 shows the time evolution of $N(t)$ for spherical proteins with a gradient distribution of charges with $e_{prot} = -2.4 e_{DNA}$ and $e_{max} = -1.2 e_{DNA}$, as well as two distributions obtained by random permutations of these charges (but the search site remains the bead with charge $e_{max}$). It can be checked on this example that the regular and random charge distributions lead essentially to the same behaviour for $N(t)$. A related question is that of the importance of the charge carried by the search site. It is remembered that it was assumed in all simulations discussed up to now that the search site is the bead with largest positive charge $e_{max}$. However, results are not much affected when this condition is released. For example, the time evolution of $N(t)$ for spherical proteins with identical gradient distributions of charges with $e_{prot} = -1.2 e_{DNA}$ and $e_{max} = -1.2 e_{DNA}$ but search sites located either on bead 1 (with charge $e_{max} = -1.2 e_{DNA}$) or bead 2 (with charge $-0.467 e_{DNA}$) are compared in Fig. 10. It is seen that the difference between the two curves is not significant. By combining the two later observations, it can be surmised that the results should be rather similar for a given set of protein charges, whatever the exact spatial distribution of the charges and the precise charge carried by the search site. It can be checked in Fig. 11 that this is indeed the case. This figure shows the time evolution of $N(t)$ for linear proteins with a gradient distribution of charges with $e_{prot} = -2.4 e_{DNA}$ and $e_{max} = -1.2 e_{DNA}$ (solid line), as well as two distributions obtained by random permutations of these charges. The search site is the central (seventh) bead of each chain. It has charge $0.13 e_{DNA}$ for the gradient distribution, and charges $-0.53 e_{DNA}$ and $0.40 e_{DNA}$ for the random distributions. In spite of the large differences between these proteins, the evolution of $N(t)$ is essentially similar for the three of them. Conclusion therefore is that, within the validity of this coarse grained model, the dynamics of DNA sampling is essentially governed by the total charge of the protein or, in the case this charge is small, by the maximum local charge, but that the exact spatial distribution of charges and the precise charge carried by the search site play little role. It can of course not be excluded that this conclusion will be somewhat moderated when the dynamics of finer grained models is investigated.

In contrast, it should be mentioned that a factor that certainly does play an important role is the accessibility of the protein search site. For example, it is clear that, for linear proteins, beads located at the extremities of the chain are more accessible and have a higher probability to interact with DNA than beads located inside the chain, so that one expects DNA sampling by the former ones to be more efficient. This is confirmed by the examination of Fig. 10, which displays the time evolution of $N(t)$ for linear proteins with identical uniform charge distributions with total charge $e_{\text{prot}} = -1.3 e_{\text{DNA}}$, but with search sites placed either on bead 1 (extremity) or bead 7 (central bead). It is seen that bead 1 samples DNA at a speed about 50% larger than the central bead. This conclusion obviously agrees with the observation that, in real life, "reading heads" are usually exposed outside the proteins, like the two α helices of the cro repressor, which can be inserted in the major or minor grooves of the DNA double helix [43].

**7 – Discussion and conclusion**

To summarize, we improved the molecular mechanical model, which we recently proposed for non-specific DNA-protein interactions [20], by describing proteins as sets of interconnected beads instead of single ones. It must first be emphasized that most results obtained with the improved model agree with (i) experimental results, (ii) the assumptions and results of kinetic models, (iii) the results obtained with our previous model. More precisely, the improved model predicts, like our original one, that DNA sampling proceeds via a succession of 3D motion in the cell, 1D sliding along the DNA sequence, short or long hops between neighboring or more widely separated sites, and intersegmental transfers. This behavior is confirmed by recent experiments [10-19] and is one the key *assumptions* of kinetic models. Quite interestingly, this behavior is however not an assumption for molecular mechanical models but rather a *consequence* of the form of DNA-protein interactions and evolution equations. In some sense, molecular mechanical models therefore provide a theoretical grounding for the assumption plugged in kinetic models. The second crucial assumption of kinetic models is that both 3D motion of the protein in the cell and 1D sliding along DNA are diffusive processes. In our previous work [20], we observed that the number $N(t)$ of different DNA beads visited by proteins increases as the square root of time during 1D sliding events and linearly with time for the global 3D motion. We concluded from this observation that 1D sliding is diffusive but not 3D motion. In the present study, we again obtained that, for the improved model, $N(t)$ increases linearly with time during 3D motion of the protein. On the basis of the expression of the volume of the 3D Wiener sausage, we however showed that the correct interpretation of these results is that 3D motion is *indeed diffusive*. Stated in other words, the results obtained with both our original and the improved model agree with the assumption of kinetic models and ground it to some extent. On the other hand, we observed that 1D sliding is slightly subdiffusive for the improved model, with an exponent $\beta \approx 0.40$ for realistic protein charges, while it was found to be diffusive ($\beta = 0.50$) for single bead proteins and it is *assumed* to be diffusive in kinetic models. There seems to be some experimental confirmation that protein motion might indeed be subdiffusive [40-42]. At last, it should be mentioned that the number of DNA base pairs visited during each sliding event (from 75 to 150 base pairs) is in qualitative agreement with both experimental results [18,19] and the values usually derived from kinetic models [5].

We used the improved model to investigate several aspects of the dynamics of DNA sampling that were accounted for neither by our original model nor by kinetic ones. For example, we showed that, within the validity of this model, the shape and deformability of proteins essentially affect the speed of DNA sampling through the associated variations of the diffusion coefficient. Moreover, it appears that the sampling speed is governed by the total charge of the protein or, in the case this charge is small, by the maximum local charge, while the exact spatial distribution of the charges and the precise charge carried by the search site seem to play only a minor role. Simulations also confirm that the accessibility of the protein search site is a key factor.

The only point for which results of molecular mechanical models differ substantially from those of kinetic ones therefore deals with the efficiency of the facilitated diffusion process, in the sense that mechanical models predict a quite low efficiency of facilitated diffusion. In our models, the relative proportions of 1D sliding and 3D diffusion experienced by proteins can indeed be adjusted by varying the protein charge distribution within physically reasonable limits. We found that for single bead proteins it is possible to increase the DNA sampling rate by only about 70% compared to the 3D diffusion limit upon variation of the 1D/3D motion ratio [20]. In the present work we obtained that this maximum increase is more likely smaller than 50% when the protein is modeled in a less crude way, that is as a set of 13 interconnected beads, because of the relative inefficiency of the 1D sliding motion of deformable proteins compared to single beads. This is in clear disagreement with kinetic models, which suggest that the speed of DNA sampling can be increased by several orders of magnitude upon variation of the ratio of 1D and 3D motions [2-8]. The question that arises is obviously to determine which of mechanical and kinetic models are in best agreement with experimental results. It has long been claimed that the results of Riggs, Bourgeois and Cohn on the LacI repressor of *Escherichia coli* are a proof that binding to specific DNA sites can surpass by several orders of magnitude the maximal rate for 3D diffusion, in spite of the fact that Riggs, Bourgeois and Cohn themselves explained that the extremely fast reaction rate they measured is probably due to the fact that their experiments were carried out at very low ionic strength, so that "there is an electrostatic attraction between a positively charged site on the repressor and the negatively charged phosphate groups in the operator" [1]. Stated in other words, at low salt concentrations, the diffusion limit is precisely of the order of $10^{10}$ $M^{-1}$ $s^{-1}$ and the measurements of Riggs, Bourgeois and Cohn just reflects this fact. The hypothesis of Riggs, Bourgeois and Cohn was confirmed by subsequent studies of the effect of salt on the kinetics of the binding of the LacI repressor, which reported a logarithmic decrease of the association rate constant from $10^{10}$ $M^{-1}$ $s^{-1}$ in the absence of salt to the expected $10^{8}$ $M^{-1}$ $s^{-1}$ value at usual salt concentrations [47-49] (see also [50-52]). Moreover, most of the DNA binding proteins that were investigated at usual salt concentrations since that time were found to have association rate constants close to and not larger than the diffusion limit [53-59]. To summarize, it can be stated, as in [60], that "no protein that binds to a specific DNA site is known to do so at a rate that exceeds the diffusion limit" and that measured rate constants that exceed the usual $10^{8}$ $M^{-1}$ $s^{-1}$ value just reflect the evolution of the diffusion limit with salt concentration. The fact that our molecular mechanical models predict that facilitated diffusion cannot substantially increase the speed of DNA sampling compared to pure 3D diffusion therefore appears to agree with experimental results.

The model proposed here describes DNA-protein non-specific interactions better than our original one, but it still deserves improvement with respect to several aspects. For example, a more realistic description should take into account the presence of histones and the fact that not all the DNA in a cell is accessible to proteins. Moreover, the description of the DNA chain should be more detailed, by taking major and minor grooves and bubbles into consideration. Last but not least, we should allow for heterogeneity on the DNA charges as a first step towards the modeling of *specific* interactions.

**Acknowledgements** : We thank Dr Nils Becker (ENS Lyon) for bringing the results on the volume of the Wiener sausage to our attention and for very helpful discussions.

**FIGURE CAPTIONS**

**Figure 1** : (color online) Time evolution of the logarithm of $1-N(t)/2000$, the portion of DNA beads not yet visited by the protein search site, for (a) linear proteins with a gradient distribution of charges with total charge $e_{\text{prot}}=0$ and maximum charge $e_{\max}=-0.8e_{\text{DNA}}$ (solid line), (b) linear proteins with a gradient distribution of charges with total charge $e_{\text{prot}}=-1.2e_{\text{DNA}}$ and maximum charge $e_{\max}=-1.2e_{\text{DNA}}$ (short dashes), (c) spherical proteins with a gradient distribution of charges with total charge $e_{\text{prot}}=0$ and maximum charge $e_{\max}=-1.5e_{\text{DNA}}$ (dot-dot-dot-dashes), and (d) spherical proteins with a gradient distribution of charges with total charge $e_{\text{prot}}=-1.2e_{\text{DNA}}$ and maximum positive charge $e_{\max}=-0.8e_{\text{DNA}}$ (long dashes). For all proteins, the search site was assumed to be the bead with charge $e_{\max}$. For the linear proteins, the search site is located at one of the extremities of the chain. It was considered that protein bead $p$ is attached to bead $k$ of DNA segment $j$ if $\|\mathbf{r}_{j,k}-\mathbf{R}_p\|\leq\sigma$. The dot-dashed straight lines, which were adjusted against the evolution of $1-N(t)/2000$ for each protein, were used to estimate the values of κ.

**Figure 2** : (color online) Log-log plots of the time evolution of the number $N(t)$ of different DNA beads visited by the protein for spherical proteins with (a) uniform charge distributions and four values of the total charge ranging from $e_{\text{prot}}=-0.8e_{\text{DNA}}$ to $e_{\text{prot}}=-4.8e_{\text{DNA}}$ (top), (b) gradient distributions of charges with total charge $e_{\text{prot}}=0$ and four values of the maximum charge ranging from $e_{\max}=-0.4e_{\text{DNA}}$ to $e_{\max}=-3e_{\text{DNA}}$ (middle), and (c) a gradient distribution of charges with total charge $e_{\text{prot}}=-0.8e_{\text{DNA}}$ and maximum charge $e_{\max}=-1.2e_{\text{DNA}}$, and four values of the elastic constant $C$ ranging from 10 to 200 (bottom). In order to improve the signal/noise ratio, it was assumed for this plot that the protein is attached to bead $k$ of DNA segment $j$ if *any* of the protein beads (and not a given one) satisfies the condition $\|\mathbf{r}_{j,k} - \mathbf{R}_p\| \leq \sigma$. Each curve was averaged over a number of sliding events that varied between 50 and 200. Each sliding event lasted more than 1 µs, during which the protein neither separated from the DNA segment by more than σ during more than 0.07 µs nor reached one of the extremities of the DNA segment.

**Figure 3** : (color online) Plot, as a function of the total protein charge $e_{\text{prot}}$, of $N(100 \mu s)$ (top), $\rho_{\text{1D}}$ (middle), and $n_{\text{sim}}$ (bottom) for spherical proteins with uniform charge distributions. $N(100 \mu s)$ is the number of different DNA beads visited by the protein search site after 100 µs, $\rho_{\text{1D}}$ is the portion of time that the protein search site spends attached to a DNA bead, and $n_{\text{sim}}$ is the average number of DNA beads that are simultaneously attached to the protein search site when it interacts with DNA. Circles correspond to results obtained by considering that protein bead $p$ is attached to bead $k$ of DNA segment $j$ if $\|\mathbf{r}_{j,k} - \mathbf{R}_p\| \leq \sigma$, while lozenges correspond to the criterion $\|\mathbf{r}_{j,k} - \mathbf{R}_p\| \leq 1.5\, \sigma$. Error bars indicate the standard deviations for the six trajectories over which each point was averaged (note that error bars are masked by circles and lozenges whenever the size of these symbols is larger than the computed standard deviation). Points at $e_{\text{prot}} = 0$ denote results obtained with purely repulsive interactions between DNA and the protein.

**Figure 4** : (color online) Same as Fig. 3, but for spherical proteins with gradient distributions of charges and maximum positive charge $e_{max} = -0.8 e_{DNA}$. The search site is assumed to be the protein bead with charge $e_{max}$.

**Figure 5** : (color online) Same as Fig. 3, but for spherical proteins with gradient distributions of charges and total charge $e_{prot} = 0$. The search site is assumed to be the protein bead with charge $e_{max}$.

**Figure 6** : (color online) Same as Fig. 3, but for linear proteins with uniform charge distributions. The search site is assumed to be one of the beads located at the extremities of the protein chain.

**Figure 7** : (color online) Time evolution of the number $N(t)$ of different DNA beads visited by the protein search site for spherical proteins with a gradient distribution of charges with $e_{prot} = -0.8 e_{DNA}$ and $e_{max} = -1.2 e_{DNA}$, and five values of the elastic constant $C$ ranging from 10 to 225. The value of $C$ is indicated for each curve. It was considered that protein bead $p$ is attached to bead $k$ of DNA segment $j$ if $\|\mathbf{r}_{j,k} - \mathbf{R}_p\| \leq \sigma$.

**Figure 8** : (color online) Evolution, as a function of the value of the elastic constant $C$, of (a) $D_{3D}$, the 3D diffusion coefficient at 298 K of spherical proteins with a gradient distribution of charges (with total charge $e_{prot} = -0.8 e_{DNA}$ and maximum charge $e_{max} = -1.2 e_{DNA}$) (top plot), and (b) the average value of $L_{0,j} / a_{prot}$ for these proteins, obtained from simulations with (filled circles) and without (empty squares) hydrodynamic

interactions. $L_{0,j}$ is the distance between the central bead with index 0 and the bead with index $j > 0$ initially located at one of the vertices of the icosahedron.

**Figure 9** : (color online) Time evolution of the number $N(t)$ of different DNA beads visited by the protein search site for spherical proteins with a gradient distribution of charges with total charge $e_{prot} = -2.4e_{DNA}$ and maximum charge $e_{max} = -1.2e_{DNA}$ (solid line), as well as two distributions obtained by random permutations of these charges. Shown in the small inserts are the positions of the charges at equilibrium. The darkest disk corresponds to charge $e_{max} = -1.2e_{DNA}$ and the brightest one to the maximum negative charge $0.8e_{DNA}$. The search site is the protein bead with charge $e_{max}$. It was considered that protein bead $p$ is attached to bead $k$ of DNA segment $j$ if $\|\mathbf{r}_{j,k} - \mathbf{R}_p\| \leq \sigma$.

**Figure 10** : (color online) Time evolution of the number $N(t)$ of different DNA beads visited by the protein search site for linear proteins with a uniform charge distribution with total charge $e_{prot} = -1.3e_{DNA}$, as well as for spherical proteins with a gradient distribution of charges with total charge $e_{prot} = -1.2e_{DNA}$ and maximum charge $e_{max} = -1.2e_{DNA}$. For the linear proteins, the search site (SS) is assumed to be either the first or the seventh (middle) bead, while for the spherical proteins the SS is assumed to be either bead 1 with charge $e_{max} = -1.2e_{DNA}$ or bead 2 with charge $-0.467e_{DNA}$. It was considered that protein bead $p$ is attached to bead $k$ of DNA segment $j$ if $\|\mathbf{r}_{j,k} - \mathbf{R}_p\| \leq \sigma$.

**Figure 11** : (color online) Time evolution of the number $N(t)$ of different DNA beads visited by the protein search site for linear proteins with a gradient distribution of charges

with total charge $e_{prot} = -2.4 e_{DNA}$ and maximum charge $e_{max} = -1.2 e_{DNA}$ (solid line), as well as two distributions obtained by random permutations of these charges. Shown in the small inserts are the positions of the charges at equilibrium. Filled circles correspond to positive charges and empty ones to negative charges, the radius of each circle being proportional to the absolute value of the charge. The search site, which is surrounded by a square, is the central (seventh) bead of each chain. It has charge $0.13 e_{DNA}$ for the gradient distribution, and charges $-0.53 e_{DNA}$ and $0.40 e_{DNA}$ for the random distributions. It was considered that protein bead $p$ is attached to bead $k$ of DNA segment $j$ if $\left\| \mathbf{r}_{j,k} - \mathbf{R}_p \right\| \leq \sigma$.

Figure 1

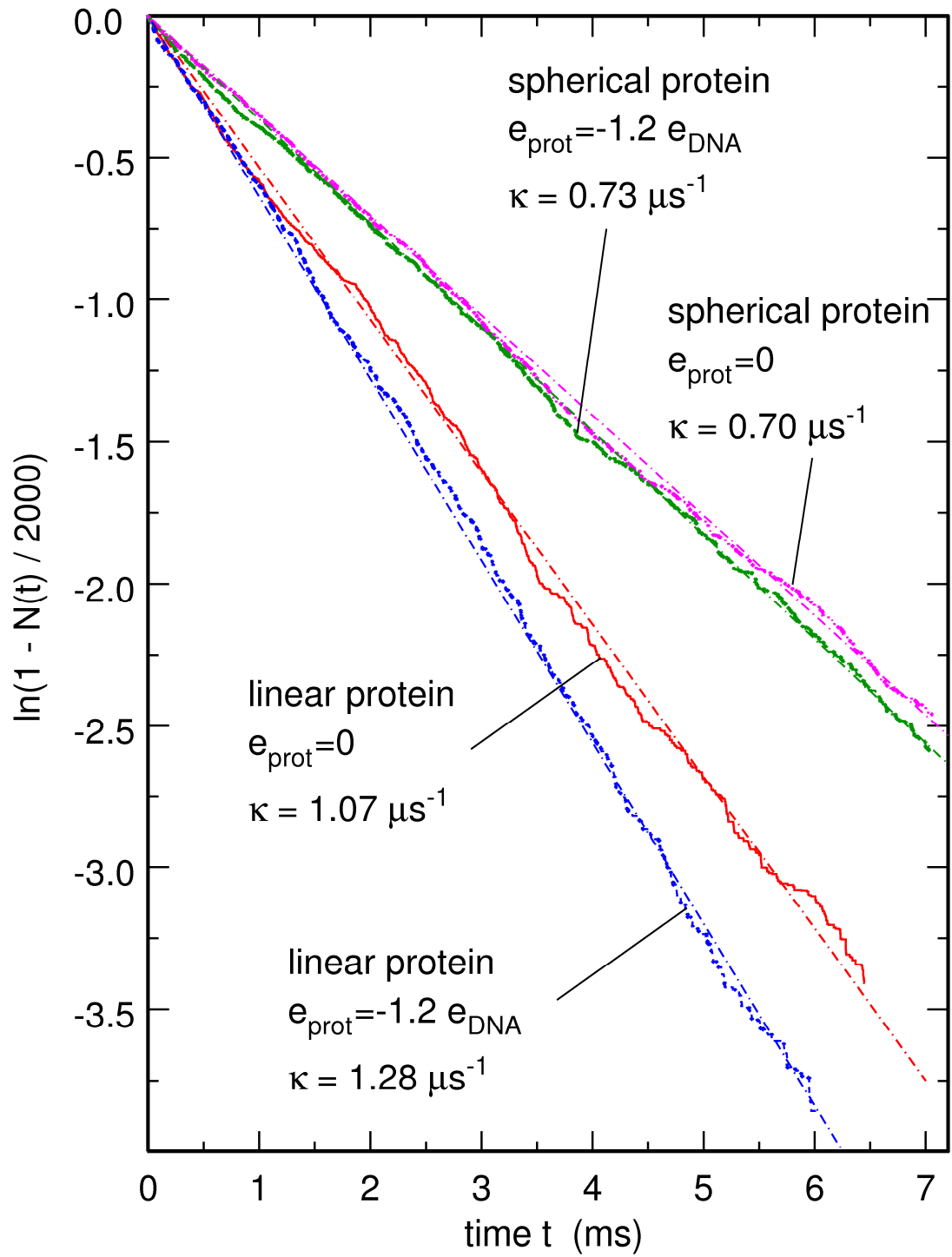

Figure 2

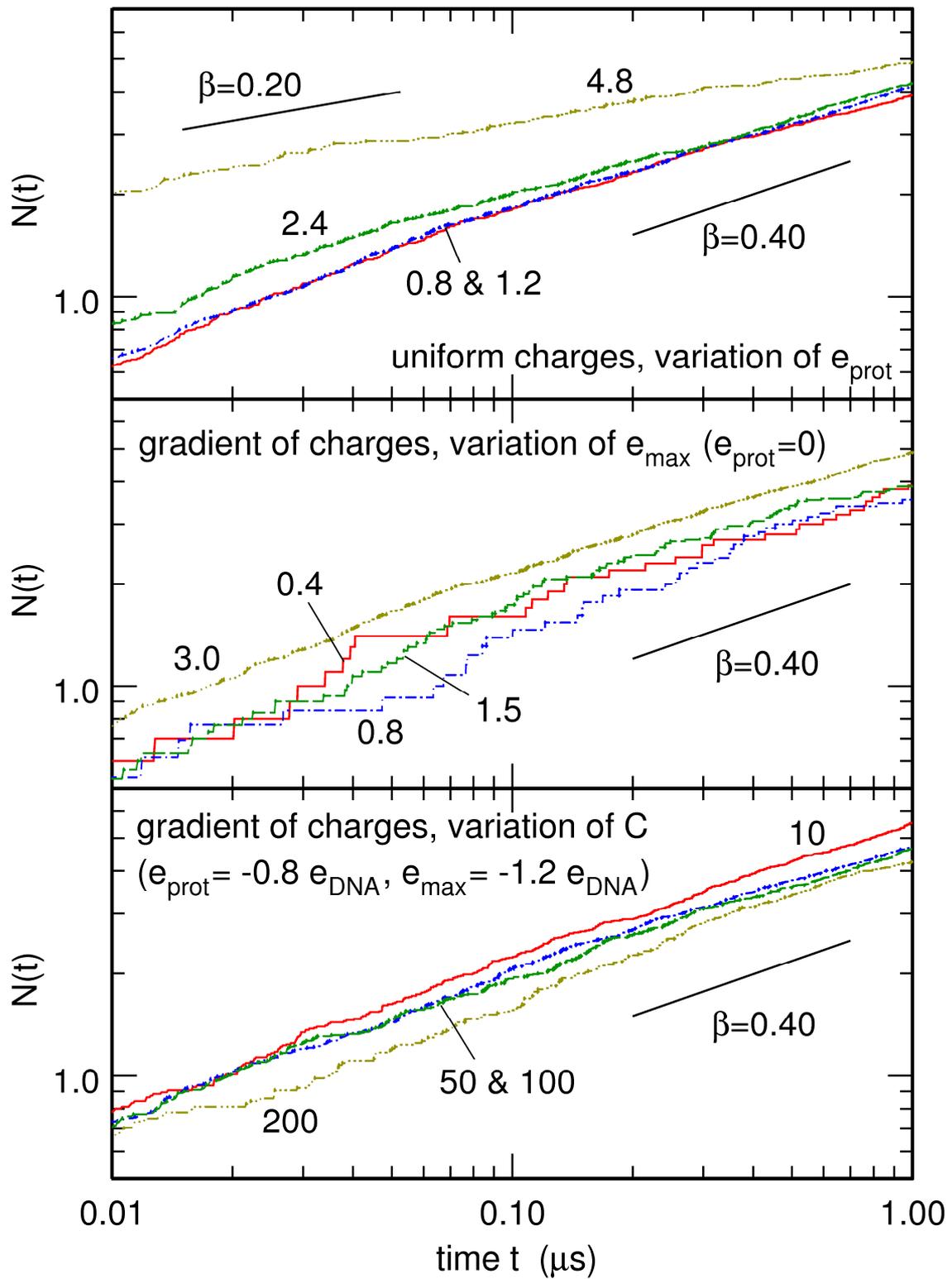

Figure 3

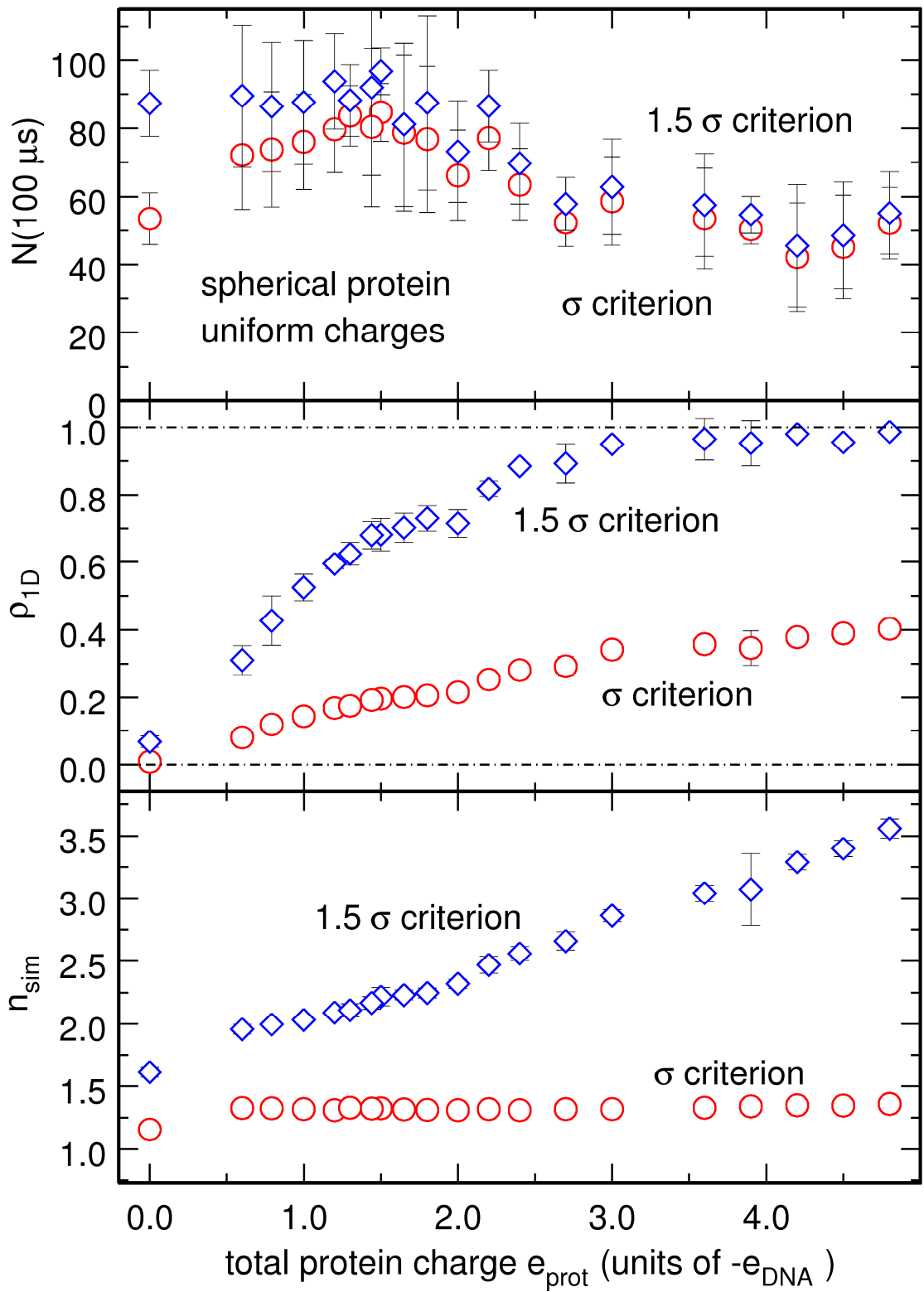

Figure 4

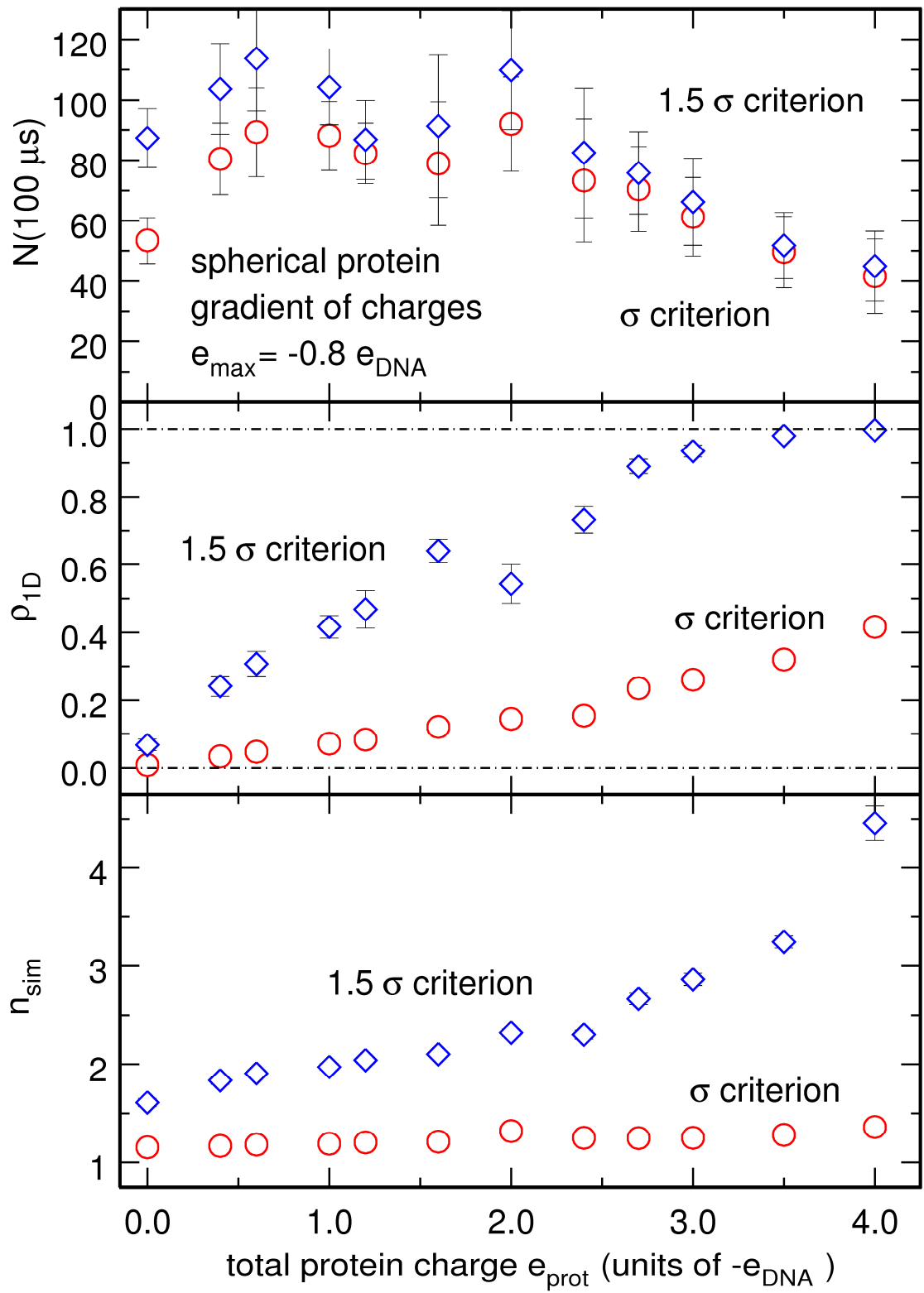

Figure 5

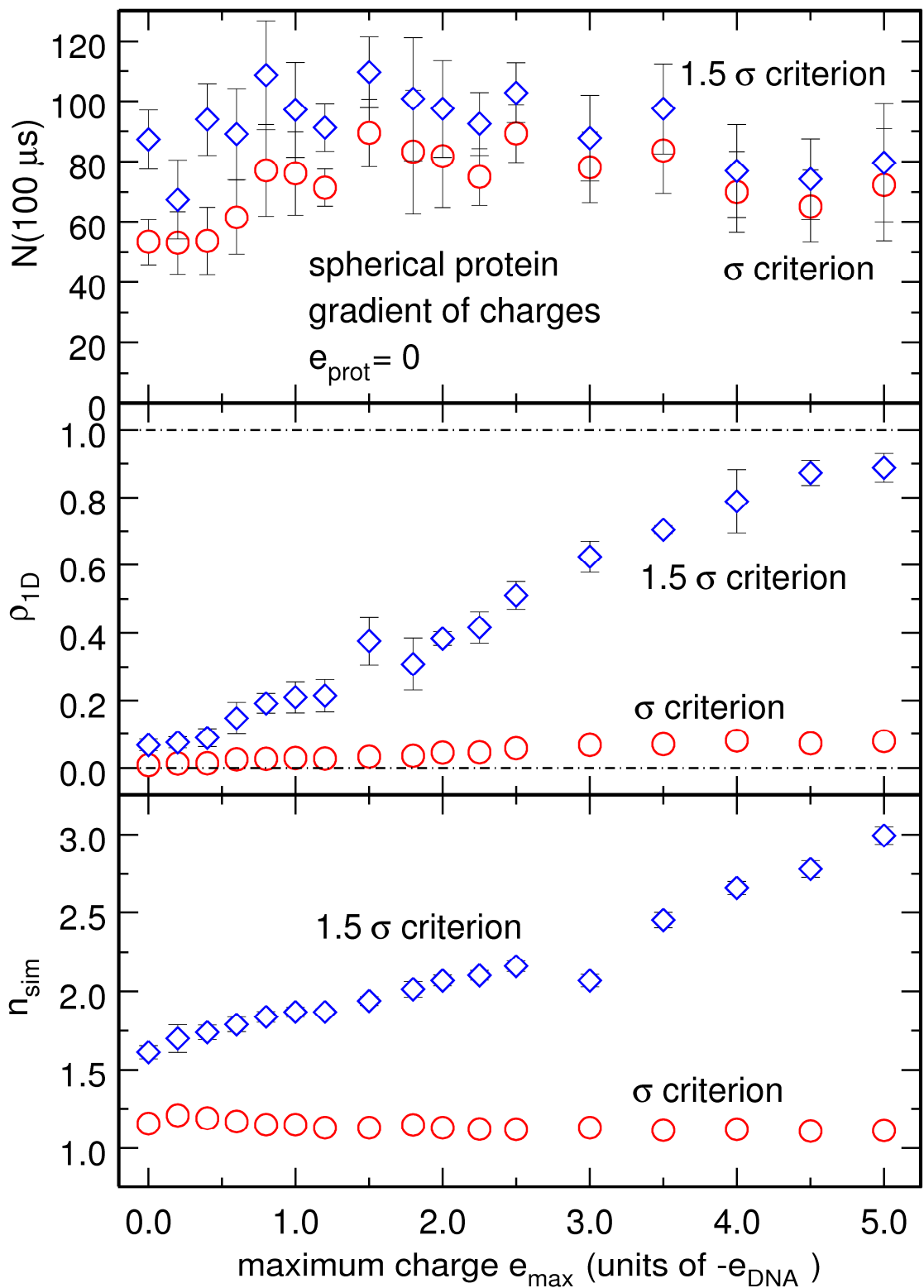

Figure 6

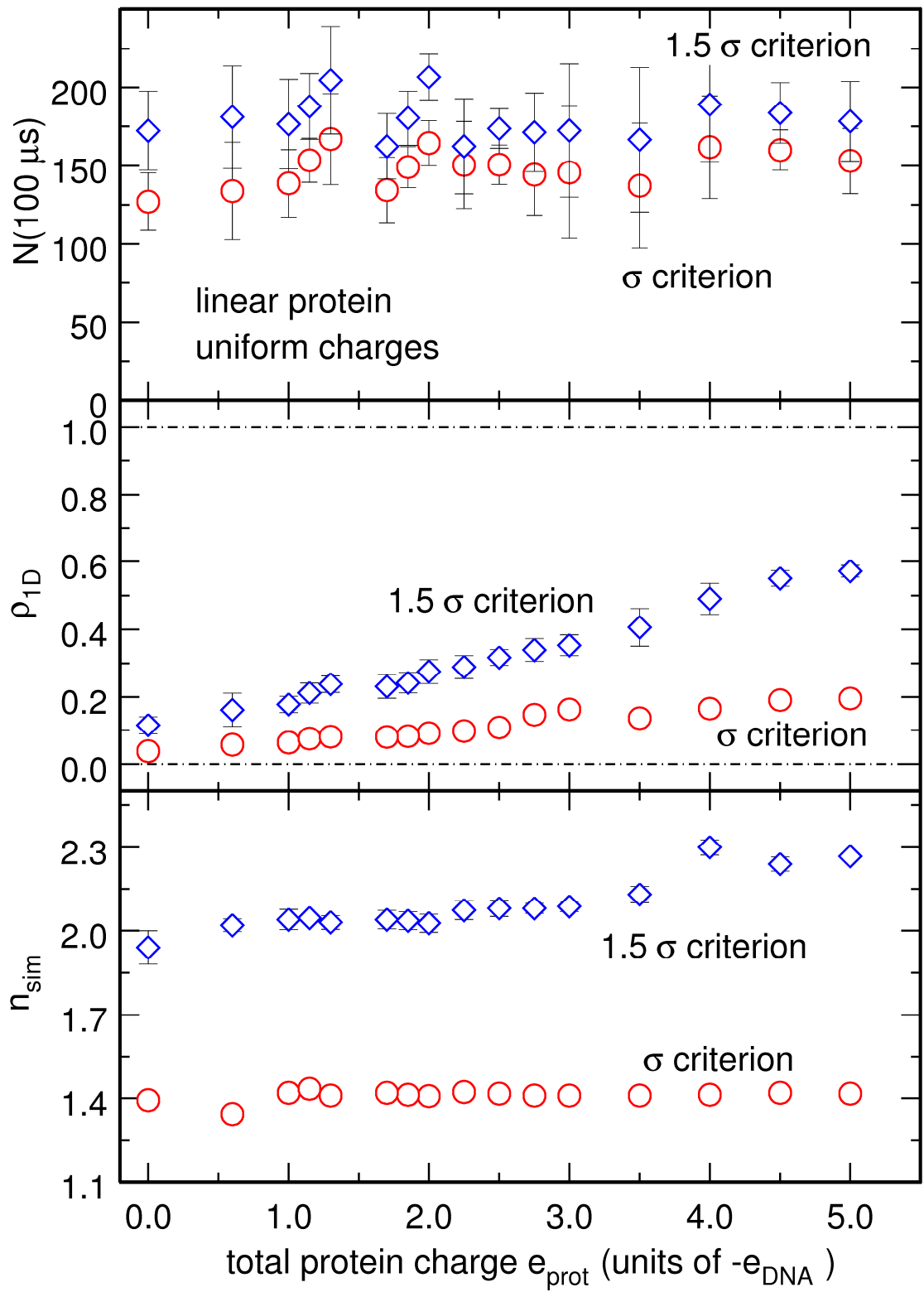

Figure 7

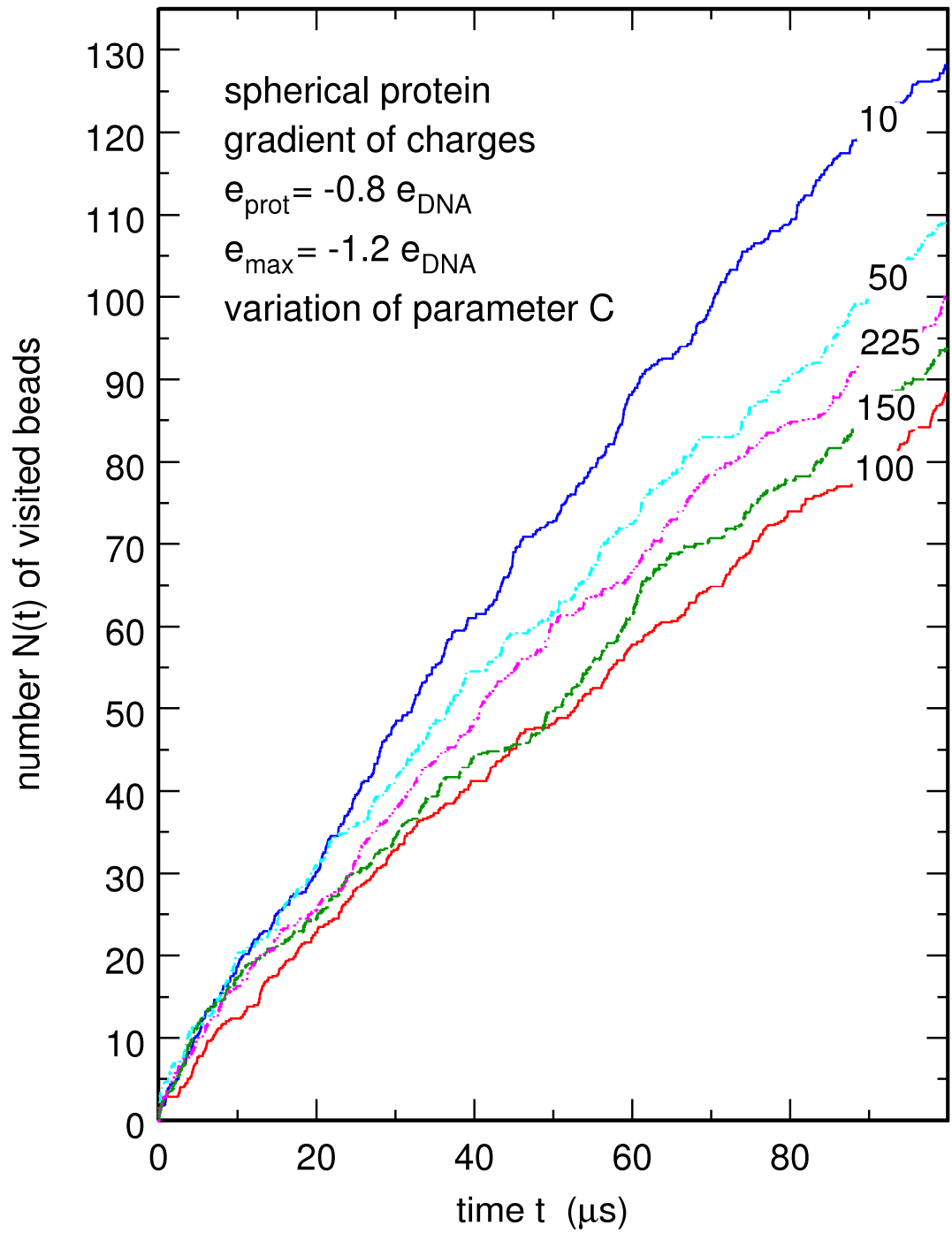

Figure 8

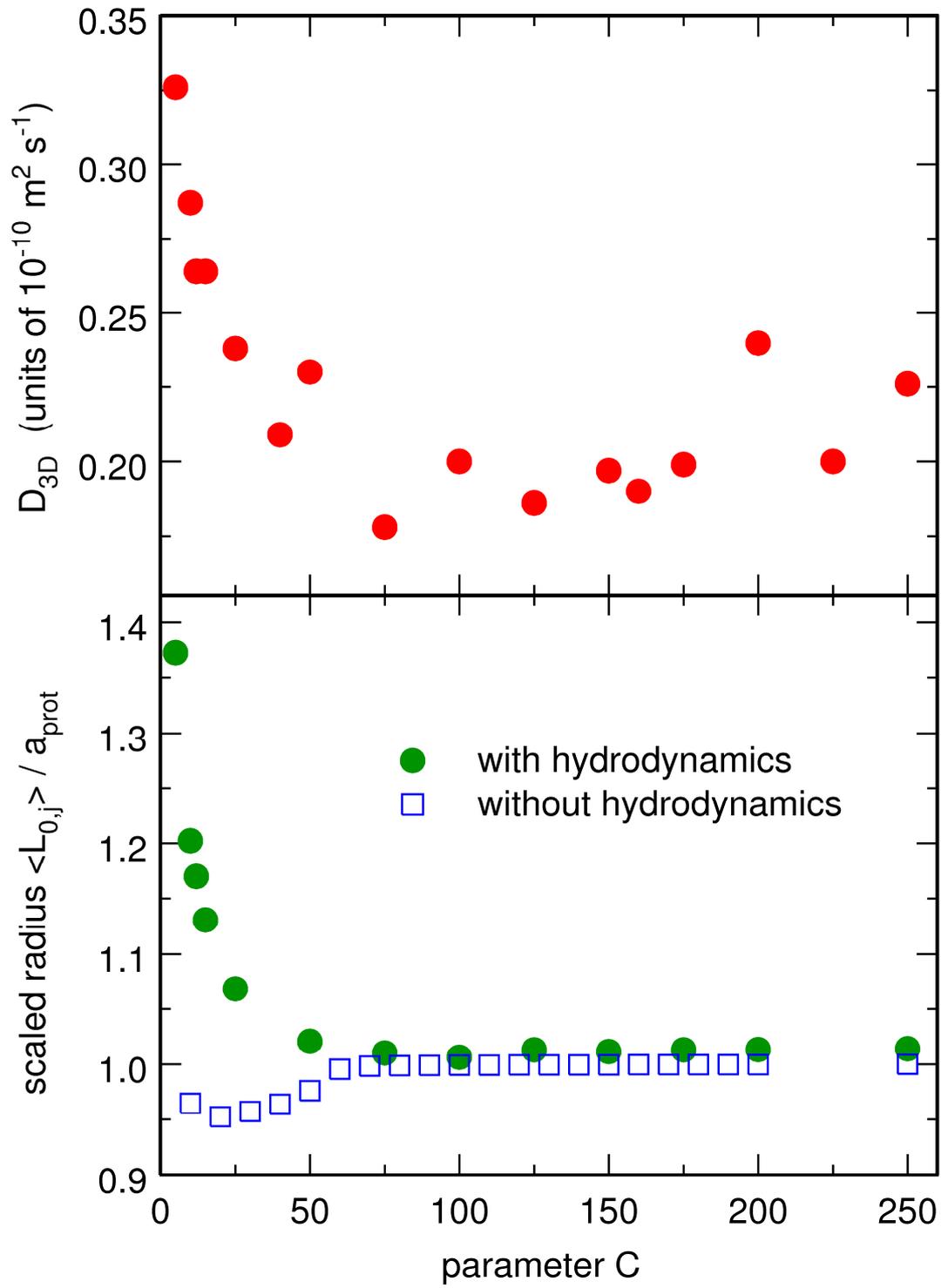

Figure 9

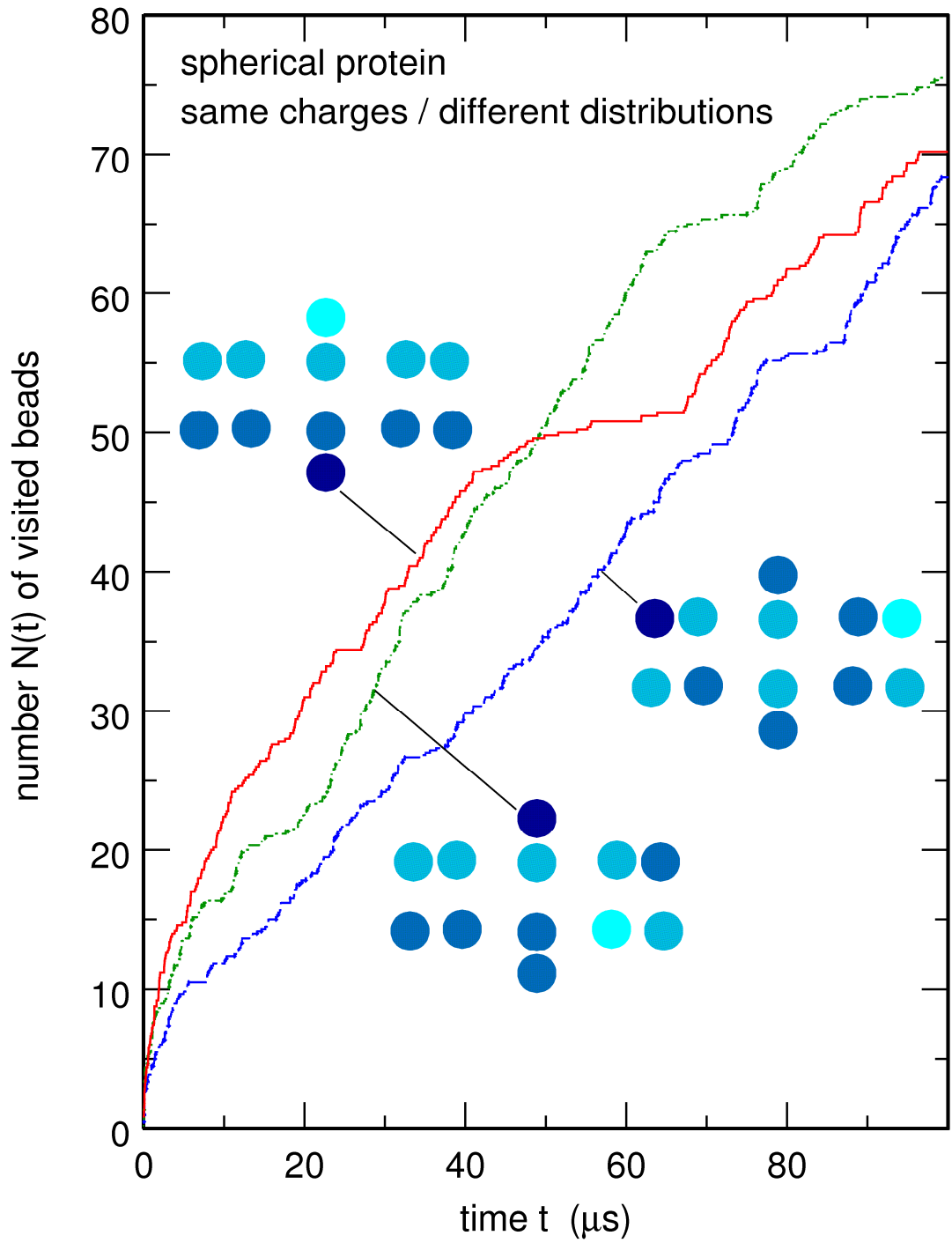

Figure 10

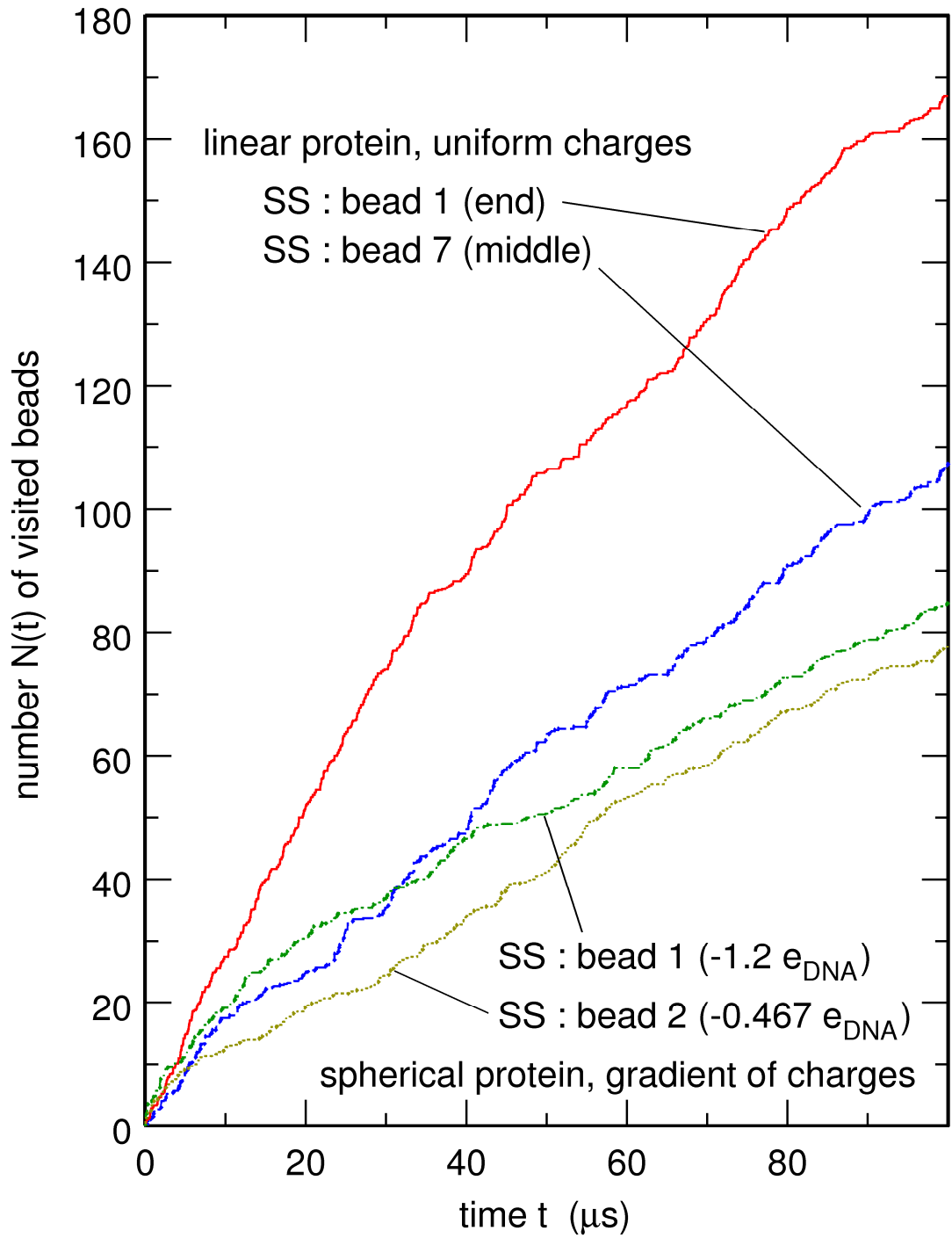

Figure 11

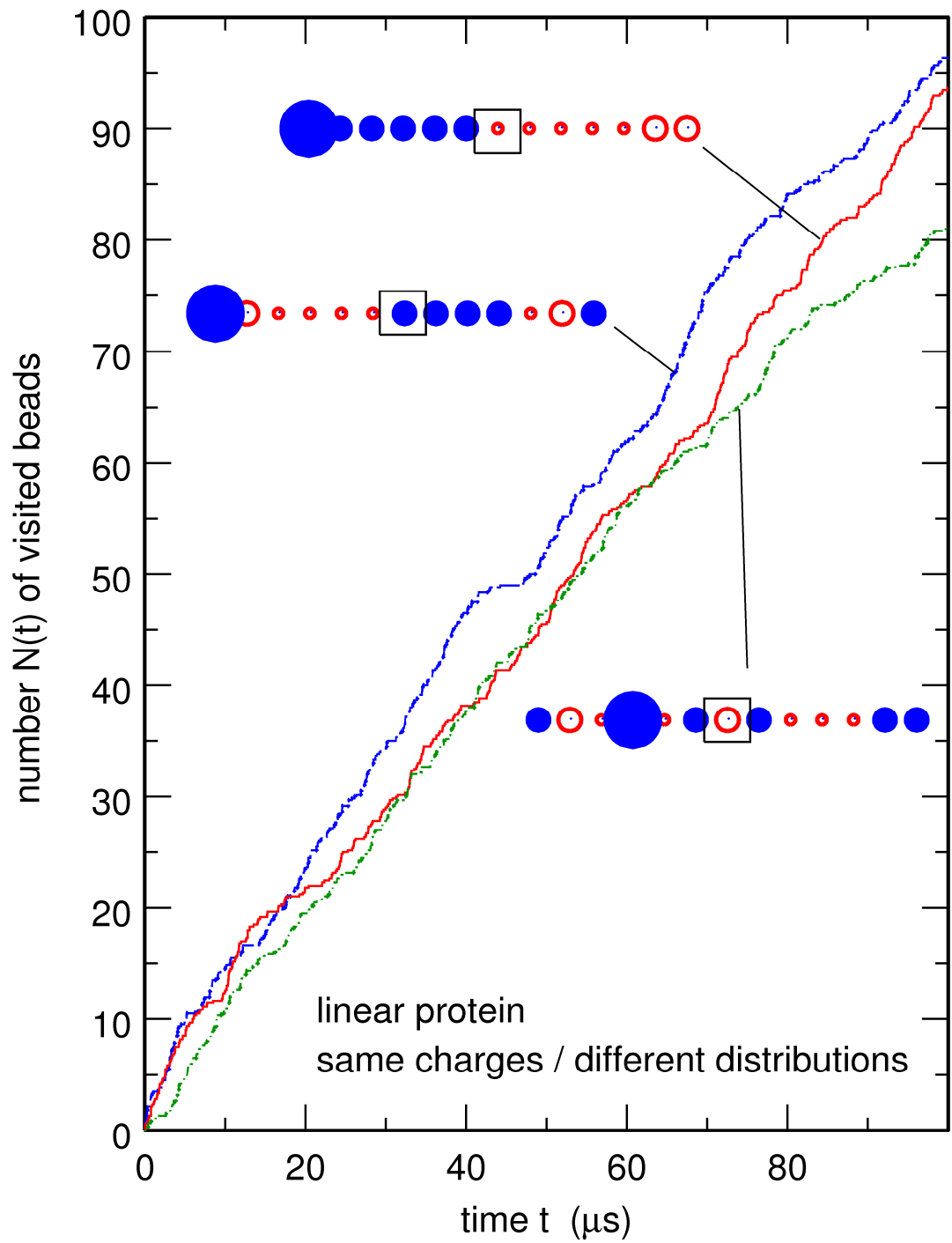